# Coexistence of magneto-resistance and -capacitance tunability in $Sm_2Ga_2Fe_2O_9$

Ye Wu

The University of Texas at San Antonio, Department of Electrical and Computer Engineering, One UTSA Circle, San Antonio, TX 78249.

E-mail: txwuye@gmail.com

**Abstract:** we propose that charge gradient resulting in the coexisting magneto-resistance and -capacitance tunability in material systems. We have experimentally observed coexisting of tunable magneto-resistance and -capacitance in $Sm_2Ga_2Fe_2O_9$. Our model fits well with the experimental result.

Keywords: magnetoresistance  magnetocapacitance   magnetoloss

## I. INTRODUCTION

The coexistence of magnetoresistive and -capacitive (MRC) effect has been shown in several manganese based metal oxides, such as $(Bi,Tb,Y,La,Pr,Ca)MnO_3$.[1]-[6] The coupled colossal MRC effect also have been reported in $HgCr_2S_4$ [7]and perovskite oxynitrides $EuMO_2N$ (M = Nb, Ta). [8]The nature of coupled MRC effect is a subject of ongoing discussion. Two mechanisms have been proposed so far. The first one is shown to be intrinsic multiferroism,[1]-[6] which involves with magnetoelectric coupling. The second one is the combination of magnetoresistance effect with the Maxwell-Wagner effect which could be used to explain the magnetocapacitance response in materials that are not multiferroic. It proposes that when a magnetic field is applied, the increase of bulk conductivity would produce an interfacial polarization in charge depleted grain boundaries. [9][10] However, the first mechanism is only practical for multiferroic materials and the second mechanism is only practical for non-multiferroic materials. A more general mechanism which could be applied to all material is still absent. In this letter, we propose a general and simple mechanism based on charge gradient that could be used to explain the coexisting magneto-resistance and -capacitance tunability in all the material systems. We have experimentally observed coexisting of tunable magneto-resistance and –capacitance in $Sm_2Ga_2Fe_2O_9$. Our theoretical model matches well with the experimental result.

## II. Theory.

In the following, we will derive the expression for magneto-resistance and -capacitance in the material system, considering the analysis of the force balance of the material in a magnetic field. If the magnetic field is supposed to be along the z-direction and the velocity in the x-direction is ignored due to the equilibrium state, then the magnetic force in the y-direction which acts on charges would be calculated as $F_y = eH_z v_x$, (1)

We conjecture that there would be attractive force and repulsive force existing along with the magnetic force. While the electric field could be determined by Poisson's equation

$$d^2\Phi/dy^2 = \tau n e \tag{2},$$

where $\tau$ is the parameter that depends on the material permittivity and could be fitted by the experimental data, e is the elementary charge, n is the charge density.

The attractive force due to the electron gas pressure is derived as,

$$ed\Phi/dy = e(E_0 + \tau e \int_0^y n(\gamma)d\gamma). \tag{3}$$

Considering the energy of the electron gas is given by $E_{eg} = \hbar^2(3\pi^2 n)^{2/3}/2/m$, (4)

the repulsive force could be expressed as $\dfrac{dE_{eg}}{dy} = \dfrac{\hbar^2}{m}\pi^{4/3}3^{-1/3}n^{-1/3}\dfrac{dn}{dy}$. (5)

Hence the force balance equation could be formed as

$$\frac{\hbar^2}{m}\pi^{4/3}3^{-1/3}n^{-1/3}\frac{dn}{dy} = e(E_0 + ke\int_0^y n(\gamma)d\gamma) + eH_z v_x. \tag{6}$$

We can further correlate the local velocity of the charge $v_x$ to the charge density $n$ for solving Eq.(6). We can consider that the total magnetic field is given by

$$H_t = H_i + H \tag{7}$$

and the induced magnetic field would be expressed as $H_i = 4\pi\chi(y)H$. (8)

Using Landau diamagnetism, the susceptibility is well represented by

$$\chi(y) = \frac{-e^2 k_F}{12\pi^2 mc^2} = \frac{-e^2}{12mc^2}3^{1/3}\pi^{-4/3}n(y)^{1/3}. \tag{9}$$

Substituting the value of magnetic field into the Maxwell equation

$$\nabla \times \vec{H}_t = \partial \vec{D}/\partial t + \vec{J}, \tag{10}$$

we have $-\partial H_t/\partial y = J_x$, (11)

where we ignore the dynamic state. We conside the local current $J_x = e \cdot n \cdot v_x$, (12)

where $v_x$ is the local velocity of the charge. The velocity generated is

$$v_x(y) = \frac{-e3^{1/3}\pi^{-4/3}H_y n(y)^{-5/3}}{12mc^2}\frac{\partial n(y)}{\partial y} . \qquad (13)$$

Substituting it into the force balance equation and reorganizing it, we have

$$eE_0 + ke^2\int_0^y n(\gamma)d\gamma = \frac{e^2 3^{1/3}\pi^{-4/3}H_z^2 n(y)^{-5/3}}{12mc^2}\frac{dn(y)}{dy} + \frac{\hbar^2}{m}\pi^{4/3}3^{-1/3}n(y)^{-1/3}\frac{dn(y)}{dy} , \qquad (14)$$

Using boundary condition $\frac{dn}{dy}\big|_{y=a}=0$ and $n\big|_{y=a}=n_a$, the distance profile of the charge density which is dependent on the magnetic field strength could be expressed as

$$y(n) = \int_n^{n_a}\frac{2\xi^{-1/3}d\xi}{3\sqrt{C_1\xi^{5/3}-C_2}} - \int_n^{n_a}\frac{3\xi^{-5/3}d\xi}{2\sqrt{C_3\xi^{1/3}-C_4}}\frac{1}{H_z} , \qquad (15)$$

Where

$$C_1 = \frac{8\pi e^2 m}{15\hbar^2\pi^{4/3}3^{-1/3}} , \quad C_2 = \frac{8\pi e^2 m a^{1/3}}{\hbar^2\pi^{4/3}3^{2/3}} , \quad C_3 = \frac{2\pi e^4 3^{-5/3}\pi^{-4/3}}{mc^2} \quad \text{and}$$

$$C_4 = \frac{2\pi e^4 3^{-5/3}\pi^{-4/3}a^{1/3}}{mc^2} .$$

Taking the integral in both sides of Eq.(15), the charge density gradient could be derived as

$$\frac{\partial n}{\partial y} = \frac{-6H_x\sqrt{C_1 n^{5/3}-C_2}\sqrt{C_3 n^{1/3}-C_4}}{-4H_x n^{-1/3}\sqrt{C_3 n^{1/3}-C_4}+9n^{-5/3}\sqrt{C_1 n^{5/3}-C_2}} \qquad (16)$$

Using a more general expression, the charge density under a magnetic field could be derived as

$$\frac{\partial n}{\partial l} = \frac{-6B\sqrt{\lambda_1 n^{5/3}-\lambda_2}\sqrt{\lambda_3 n^{1/3}-\lambda_4}}{-4Bn^{-1/3}\sqrt{\lambda_3 n^{1/3}-\lambda_4}+9n^{-5/3}\sqrt{\lambda_1 n^{5/3}-\lambda_2}} \qquad (17)$$

where $l$ is the distance of electron movement under the magnetic field $\lambda_1, \lambda_2, \lambda_3$ and $\lambda_4$ are the constants which could be fitted from the experimental data, $B$ is the magnetic field, $\partial n/\partial l$ represents the charge profile under the magnetic field.

For simplicity, we also assume $\lambda_1 n^{5/3} \gg \lambda_2$ and $\lambda_3 n^{1/3} \gg \lambda_4$, therefore $\lambda_1 n^{5/3} - \lambda_2 \approx \lambda_1 n^{5/3}$, and $\lambda_3 n^{1/3} - \lambda_4 \approx \lambda_3 n^{1/3}$.

Substituting the well-known equations of conductivity and resistivity, $\sigma = qvn$, $\sigma = 1/\rho$ and $\rho = R\frac{A}{\ell}$, we can rewrite Eq.(17) as 

$$\frac{1}{R^2 qv}\frac{\partial R}{\partial l} = \frac{B\sqrt{C_1 C_3}(Rqv)^{-7/6}}{\sqrt{C_1}(Rqv)^{2/3} - B\sqrt{C_3}} . \qquad (18)$$

Here the constants of $A, \ell$ are absorbed into the constants of $C_1$ and $C_3$. The solution of $\rho$

with respect to $B$ in Eq.(18) could involve with the quintic function. We derive the simplest solution as $R = \kappa_0 B^{3/2} + \kappa_1$, (19) where $\kappa_0$ and $\kappa_1$ are the constants.

The polarization could be generated from Eq.(17) as

$$P = \frac{C_0 H \sqrt{(C_1 n^{5/3} - C_2)(C_3 n^{1/3} - C_4)}}{n^{-5/3}\sqrt{C_1 n^{5/3} - C_2} - 4n^{-1/3} H \sqrt{C_3 n^{1/3} - C_4}} \qquad (20)$$

where $C_0, C_1, C_2, C_3, C_4$ are the constants that could be fitted from the experimental data.

The capacitance could be derived as

$$C = \tau_1 \frac{H\sqrt{(C_1 n^{5/3} - C_2)(C_3 n^{1/3} - C_4)}}{n^{-5/3}\sqrt{C_1 n^{5/3} - C_2} - 4n^{-1/3} H \sqrt{C_3 n^{1/3} - C_4}} + \tau_2 (E - C_\omega \sum C_{1,\beta} H^\beta) \qquad (21)$$

It should be noted that the first term in Eq.(21) could be directly derived from Eq.(20), which represents the capacitance component that is generated from the charge gradient. The second term in Eq.(21) refers to possible magnetoelectric coupling in material system, which could be derived using classical free energy model ( as is shown in Appendix A). Here $\tau_1$ and $\tau_2$ are constants. If there is no magnetoelectric coupling in the material, $\tau_2 = 0$. Eq.(19) and Eq.(20) indicate that the magnetoresistance and magnetocapacitance effect could be coexisting in the materials. Both are derived from the charge gradient in Eq.(17). This indicates that the construction of charge gradient is the key reason of forming magnetoresistance and magnetocapacitance coupling in material systems.

III. **Experimental.**

The $Sm_2Ga_2Fe_2O_9$ samples are prepared by using sol gel method. The details of the sample preparation is shown in appendix B. The phases were confirmed by XRD. The structure of $Sm_2Ga_2Fe_2O_9$ was refined and drawn by GSAS software package. Then the refined unit cell was drawn by DRAWxtl. A magnetic field whose direction was perpendicular to the sample surface, was generated by a DC magnet. The resistance was measured by a multimeter(Fluke 8846A). The Hall coefficient was measured using a standard Hall bar configuration. The capacitance and loss were measured using a LCR meter (Hp 4284A). The measurement unit was shielded and guarded to prevent any leakage current. Ellipsometer was used to determine the thickness of the film. The typical thickness of $Sm_2Ga_2Fe_2O_9$ film used was around 120nm.

IV. **Results and discussion.**

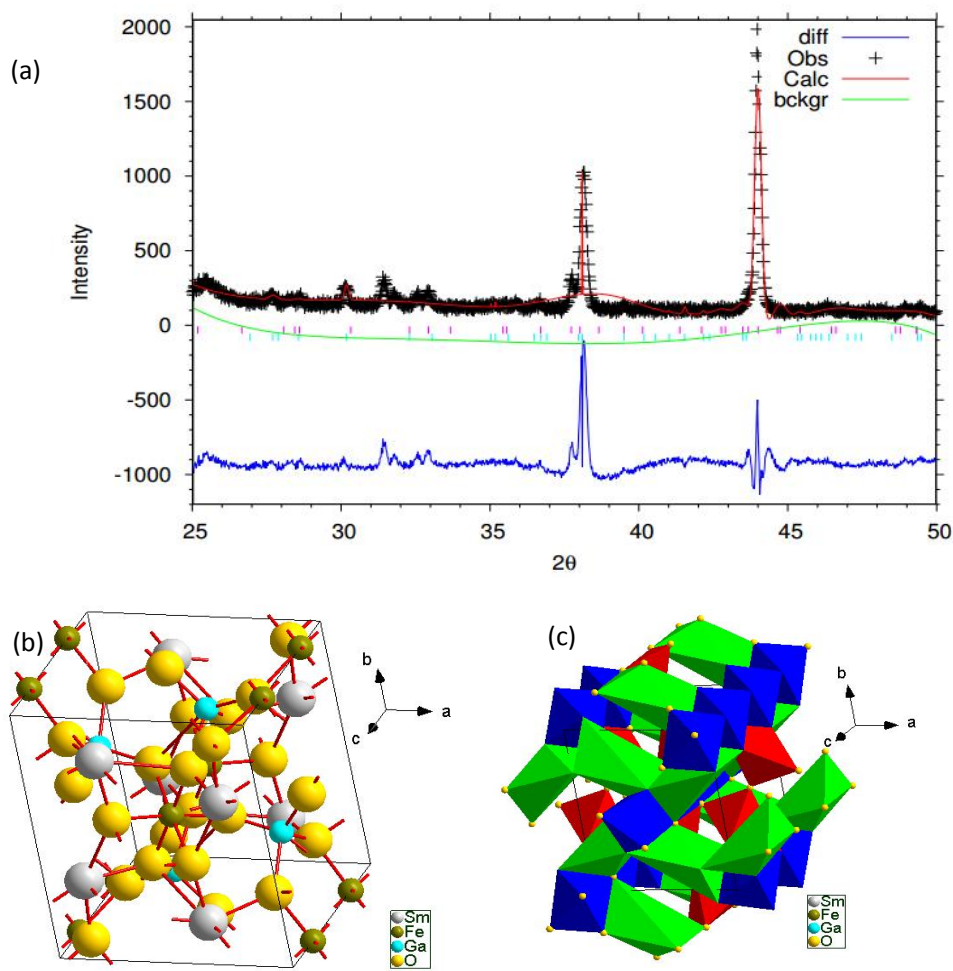

Figure 1 (a) XRD profile of $Sm_2Ga_2Fe_2O_9$ film. (Green solid line: background; red solid line: calculated XRD curve; star: experimental XRD peaks; blue solid line: difference between the calculated and experimental XRD peaks; pink/cyan lines: phase indication.)(b) the unit cell structure of $Sm_2Ga_2Fe_2O_9$. (c) polyhedrons formed by $Sm^{+3}$-$O^{-2}$, $Fe^{+3}$-$O^{-2}$, $Ga^{+3}$-$O^{-2}$

We have refined the unit cell structures from the phases of $Sm_2Ga_2Fe_2O_9$ (Fig. 1). The lattice parameters refined are as followed: Lattice constants: a=7.25039Å, b=8.577000Å, c=6.23852Å; Lattice angle: $\alpha=\beta=\gamma=90°$. Space group: Pbam. Fig.1(c) shows the polyhedrons formed by $Sm^{+3}$-$O^{-2}$, $Fe^{+3}$-$O^{-2}$, $Ga^{+3}$-$O^{-2}$. It should be noted that one $Sm^{+3}$ atom and five $O^{-2}$ atoms form a pentahedron(green);one $Fe^{+3}$ and six $O^{-2}$ atoms form an octahedron (blue);one $Ga^{+3}$ atom and five $O^{-2}$ form a hexahedron (red).

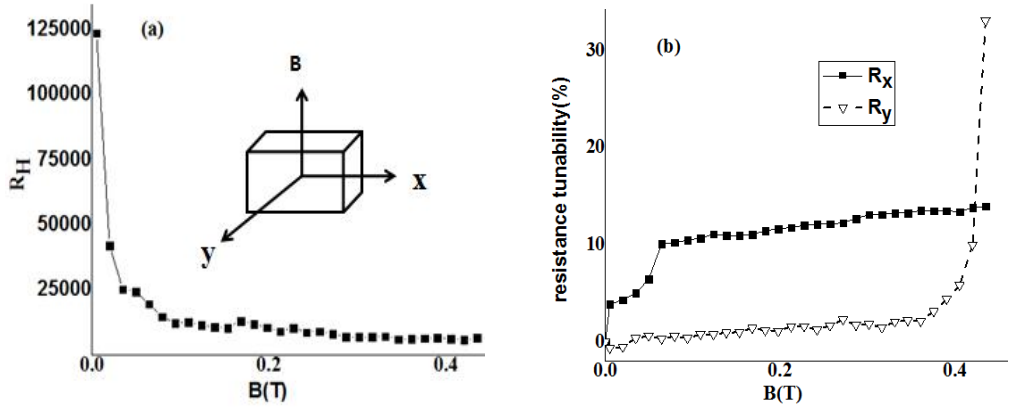

Figure 2 (a) Hall coefficient vs. magnetic field of $Sm_2Ga_2Fe_2O_9$. The inset shows the measurement and sample configuration. (b) Magnetoresistance tunability in the x and y direction.

The Hall coefficient shows a trend of decreasing with magnetic field increasing(Fig.2(a)), which means that carrier density in $Sm_2Ga_2Fe_2O_9$ samples increases with magnetic field increasing.

Fig. 2(b) compares the magnetoresistance tunability of the sample in the x- and y- direction. Here, the resistance tunability is defined as $\Delta R/R_0$ where $R_0$ is the resistance when H=0.0069T; $\Delta R$ is the variation of resistance at certain magnetic field respectively. The configuration of x- and y-direction with the magnetic field is shown in the inset of Fig. 2(b). Here $R_x$ and $R_y$ mean the resistance in the x and y direction. Up to 33% of magnetoresistance tunability could be achieved in the y-direction of the sample. It should be noted that $Sm_2Ga_2Fe_2O_9$ is deposited on a $GaFeO_3$ substrate. The lattice mismatch between $Sm_2Ga_2Fe_2O_9$ and $GaFeO_3$ may introduce varied dislocation or defects in different surface location. This may result in resistance anisotropy. As is shown in Fig.2(b), $R_y$ is bigger than $R_x$. A jump in $R_y$ appears at ~0.4 T, while $R_x$ is shown to be steadily increased with the magnetic field. The difference between $R_x$ and $R_y$ may indicate the resistance anisotropy or defects in the x or y direction.

We also measured the magnetodielectric response of $Sm_2Ga_2Fe_2O_9$. The parameter of capacitance/loss was used to represent the figure of merit for capacitance modulation(Fig.3 (c)). Here, the capacitance and loss tunability are defined as $\Delta C/C_0$ and $\Delta tan\delta/ tan\delta_0$ respectively, where $C_0$ and $tan\delta_0$ are the capacitance and loss tangent factor when H=0.0069T; $\Delta C$ and $\Delta tan\delta$ are the variation of the capacitance and loss tangent factor at certain magnetic field respectively.

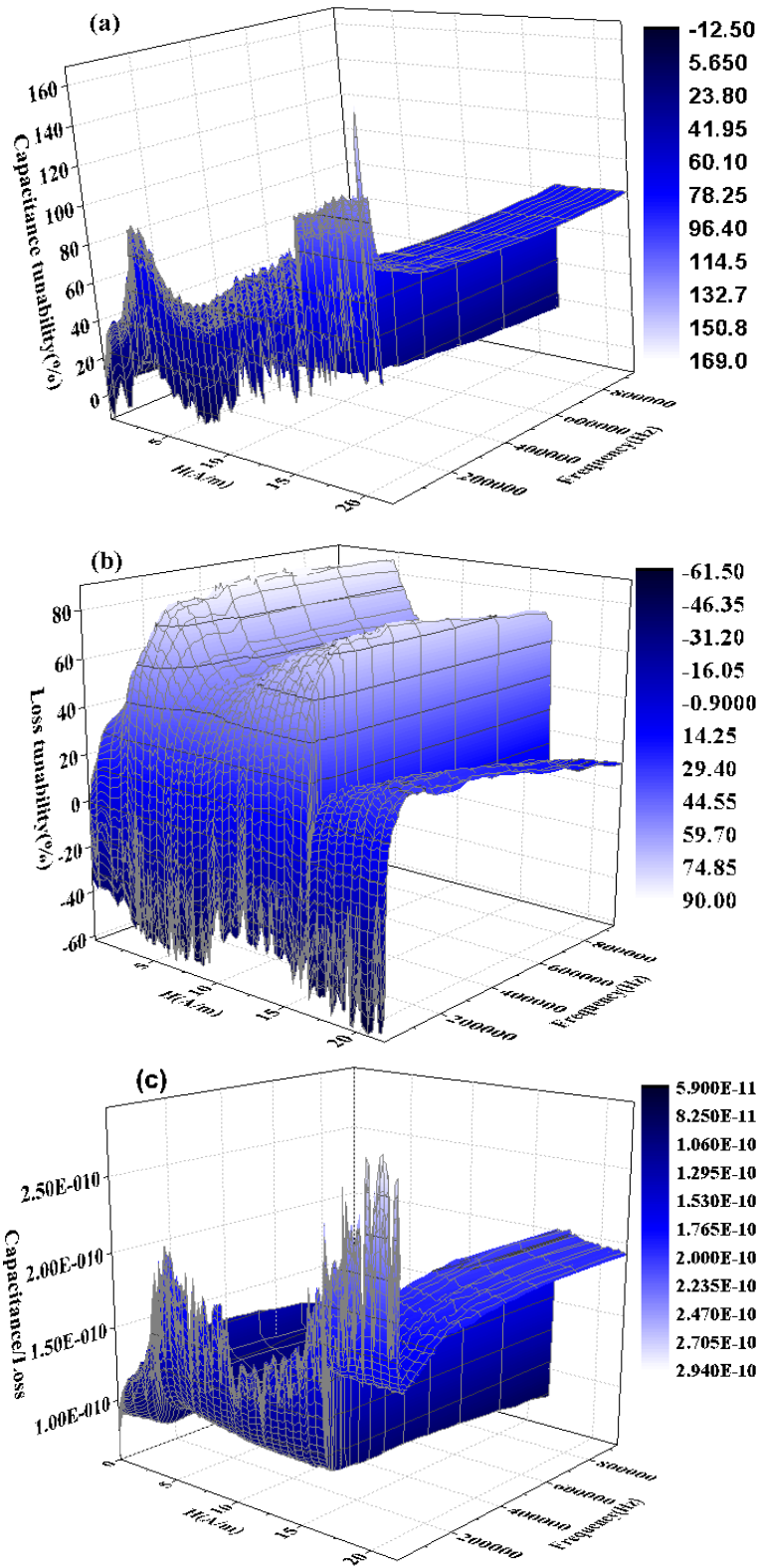

Fig.3(a) Capacitance tunability, (b) Loss tangent factor tunability, and (c) the value of Capacitance/Loss dependence of $Sm_2Ga_2Fe_2O_9$ as a function of the magnetic field and the electric frequency.

For the magnetodielectric response, we can understand the characteristics of Fig.3 as follows: (i) the capacitance and loss both depend on the magnetic field; (ii) the magnetocapacitance tunability is up to 1.6 fold and the magnetoloss tunability 0.9 fold; (iii)larger magnetocapacitance is associated with lower electric signal frequency and higher magnetic field;(iv) the tunability of capacitance and loss is sharper on the low electric signal frequency, while for the high electric signal frequency, the tunability is smooth;(v) large capacitance associated with small loss could appear in the relatively high magnetic field region(H≥16A/m).As indicated in Fig.2(b) and Fig.3(a), the magneto–resistance and –capacitance are coexisting in $Sm_2Ga_2Fe_2O_9$.

We compare the experimental and simulated results of magneto-resistance and magneto-capacitance using our previous model(Fig.3).

The magentocapacitance(Fig.4(b)) shows a steadily increasing trend. Here we arbitrarily picked up three data sets of frequency 1080Hz ,10700Hz and 941000Hz in order to compare the experimental and simulated results. It would be complex if we have to show the simulations of all the frequencies. The trend of other data with various frequencies are similar. Here we show that the simulations of magnetoresistance and magnetocapacitance fit well with trend of the experimental results in Fig.4, which suggests our model fits enormously well for such samples.

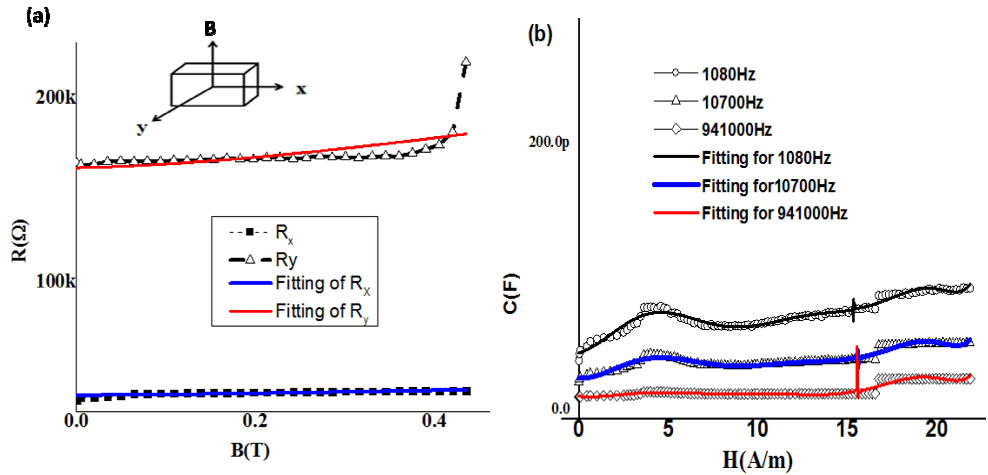

Fig.4 (a) Simulated and Experimental data of magnetoresistance in $Sm_2Ga_2Fe_2O_9$ film. The inset shows measurement configuration: the direction of magnetic field B is vertical to the film surface.(b) Simulated and Experimental data of magnetocapacitance in $Sm_2Ga_2Fe_2O_9$ film.

### V. Conclusion

We examined the effect of a magnetic field on the charge profile for the $Sm_2Ga_2Fe_2O_9$ samples. Through the use of basic electromagnetism approach, we found that the charge gradient may be depended on an external magnetic field. This charge gradient could be further derived and may be associated with magneto-resistance and -capacitance.

We expanded this coupling further and derived the tunable resistance and capacitance with the magnetic field. We have experimentally observed tunable magnetoresistance and magnetocapacitance effect in $Sm_2Ga_2Fe_2O_6$. We also observed that the Hall coefficient in Sm$_2$Ga$_2$Fe$_2$O$_6$ decreased with the magnetic field. Here, we use relationship between resistance/capacitance and magnetic field to fit the experimental data. We present a mechanism assuming charge gradient resulting in the coexistence of magnetoresistance and magnetocapacitance in some material systems. In combination with experimental observations, this mechanism shows a possible route to magnetically tune multifunctional materials and this suggests an exciting perspective way to develop new memory and sensor technologies.

**Acknowledgments.**

This work has been supported by the National Science Foundation under the Grants ECCS #1002380 and DMR #0844081.

12572–12573(2008).

[9] G. Catalan; Magnetocapacitance without magnetoelectric coupling; Appl. Phys. Lett. 88, 102902 (2006),.

[10] G. Catalan and J.F.Scott; Magnetoelectrics: Is CdCr2S4 a multiferroic relaxor? Nature 448, E4-E5 (2007).

[11] L.D. Landau, E.M. Lifshitz and L.P.Pitaevskii; Electrodynamics of continuous media, 2nd edition., Butterworth Heinenann, London (2002),pp.78.

[12] Q. X. Jia, T. M. McCleskey, A. K. Burrell, Y. Lin, G. E. Collis, H. Wang, A. D. Q. Li, and S. R. Foltyn; Polymer-assisted deposition of metal-oxide films; Nature Materials 3,529(2004).

**Appendix A**:

This appendix we show how the magnetoelectric coupling term could be derived from the free energy model.[11]

We consider that the total free energy is

$$F = F_0 + 2\pi P_2^2/\varepsilon_0 + C_0 P_2^4 - E_{eff} P_2 - E^2/8\pi + C_\omega \sum C_{\alpha,\beta} P_2^\alpha M^\beta \tag{A.1}$$

where $\alpha, \beta$ are integers, $C_\omega$ is the parameter which depends on the electric signal frequency. $C_{\alpha,\beta}$ is the constant that is associated with $\Sigma P_2^\alpha M^\beta$. The terms $\Sigma C_{\alpha,\beta} P_2^\alpha M^\beta$ indicates the interaction between polarization $P_2$ and magnetization M. For simplicity, we assume $\alpha=1$.

The energy could have its minimum when the system is in equilibrium. By taking the extreme values, we have

$$\partial F/\partial P_2 = 4 C_0 P_2^3 + 4\pi/\varepsilon_0 P_2 - E + C_\omega \sum C_{1,\beta} M^\beta = 0, \tag{A.2}$$

When $C_0=0$, we have the simple solution

$$P_2 = \frac{4\pi(E - C_\omega \sum C_{1,\beta} M^\beta)}{\varepsilon_0} \tag{A.3}$$

The magnetization is rewritten as $M = \chi_v H$, (A.4)

where $\chi_v$ is the volume magnetic susceptibility and H is magnetic field strength. For simplicity, we consider the volume magnetic susceptibility of each layer is the same.

The effective capacitance $C_2$ associated with $P_2$ is represented by

$$C_2 = \tau_2(E - C_\omega \sum C_{1,\beta} H^\beta) \tag{A.5}$$

where $\tau_2$ is a constant that can be fitted through the experimental data.

**Appendix B:**

The preparation of $Sm_2Ga_2Fe_2O_9$ films are prepared by Polymer assisted deposition method[12]. The materials used were iron nitrate nona-hydrate (Alfa Aesar), samarium acetate hydrate(Aldrich), acetic acid (Alfa Aesar) and polyethyleneimine (Alfa Aesar).

Samarium acetate hydrate (0.02g), iron nitrate nona-hydrate were dissolved in water (0.03g) to get Sm-Fe solution, which then was added with polyethyleneimine liquid solution(0.03g). The Sm-Fe stock solution was spin-coated above on a $GaFeO_3$ substrate. The coated structure was crystallized at 1100℃ for 10 hours.